\documentclass[journal,10pt,draftclsnofoot,doublecolumn]{IEEEtran}
\usepackage{mathpple}
\usepackage{times}
\usepackage{dsfont}
\usepackage{subcaption}
\usepackage{amsmath}  
\usepackage{amssymb}  
\usepackage{mathrsfs} 
\usepackage{theorem}  
\usepackage{cite}     
\usepackage{comment}  
\usepackage[colorlinks,
linkcolor=black,
anchorcolor=black,
citecolor=black
]{hyperref}
\usepackage{upref}
\usepackage{amsfonts}
\usepackage{dsfont}
\usepackage{verbatim}
\usepackage{algorithm}
\usepackage{algorithmic}
\usepackage[dvipsnames,usenames]{color}
\usepackage{enumerate}

\usepackage{graphicx}

\usepackage[flushleft]{threeparttable}
\usepackage{latexsym}

\usepackage{color}
\usepackage{footnote}
\usepackage{booktabs}
\usepackage{cite}
\usepackage{hhline}
\usepackage{gensymb}
\usepackage{graphicx}
\usepackage{xcolor}
\usepackage{listings}
\usepackage[
top    = 0.75in,
bottom = 1.05in,
left   = 0.63in,
right  = 0.63in]{geometry}
\usepackage{caption}
\usepackage{romannum}
\usepackage{setspace}

\usepackage{bm}
\usepackage{tikz}
\usetikzlibrary{external}
\usetikzlibrary{dsp}
\usetikzlibrary{graphs}
\usetikzlibrary{matrix}

\usepackage{bbm}

\usepackage{multirow}

\usepackage{makecell}

\usepackage{comment}

\usepackage{romannum}

\usetikzlibrary{external}
\usepackage{pifont}
\newcommand{\RNum}[1]{\textup{\uppercase\expandafter{\romannumeral#1\relax}}}

\newcommand{\be}[1]{\begin{equation}\label{#1}}
\newcommand{\ee}{\end{equation}}

\newcommand{\Cref}[1]{Co\-rol\-la\-ry\,\ref{#1}}

\theoremstyle{plain} \theorembodyfont{\normalfont\slshape}

\newtheorem{thm}{Theorem$\!$}

\newtheorem{prop}[thm]{Proposition$\!$}

\newtheorem{lem}[thm]{Lemma$\!$}

\newtheorem{cor}[thm]{Corollary$\!$}

\newtheorem{prob}[thm]{Problem$\!$}

\newtheorem{defi}[thm]{Definition$\!$}

\theorembodyfont{\normalfont}

\newtheorem{exam}{Example$\!$}

\newtheorem{remrk}{Remark$\!$}
\newenvironment{remark}{\begin{remrk}\hspace*{-1ex}{\bf .}}{\end{remrk}}

\definecolor{Codecolor}{named}{White}  

\begin{document}
	\title{Spatio-Temporal Modeling for Flash Memory Channels Using Conditional Generative Nets}
	
	\author{{{{\large\bf Simeng Zheng}}, {{\large\bf Chih-Hui Ho}}, {{\large\bf Wenyu Peng}}, and {{\large\bf Paul H. Siegel}} } \\
			Electrical and Computer Engineering Dept., University of California, San Diego, La Jolla, CA 92093 U.S.A \\
			{\it \{sizheng,chh279,w6peng,psiegel\}@ucsd.edu} 
			\vspace{-6.5ex}}
	
	\maketitle
	\pagenumbering{gobble}

\begin{abstract}
We propose a data-driven approach to modeling the spatio-temporal characteristics of NAND flash memory read voltages using conditional generative networks. The learned model reconstructs read voltages from an individual memory cell based on the program levels of the cell and its surrounding cells, as well as the specified program/erase (P/E) cycling time stamp. We evaluate the model over a range of time stamps using the cell read voltage distributions, the cell level error rates, and the relative frequency of errors for patterns most susceptible to inter-cell interference (ICI) effects. We conclude that the model accurately captures the  spatial and temporal features of the flash memory channel.

\end{abstract}
	
\vspace{-2ex}
\section{Introduction}
\label{sec::intro}

Realistic models for digital communication and storage channels are essential tools in the design and optimization of signal processing, detection, and coding algorithms. For NAND flash memories, the steady reduction in technology feature size and the increase in cell bit-density has been accompanied by diminished memory reliability and reduced device endurance. Sources of errors are manifold, including programming errors, inter-cell interference (ICI), cell wear during program/erase (P/E) cycling, cell charge loss due to data retention, and read disturb effects. Consequently, there is a critical need for comprehensive models that capture the complex behavior of the flash memory channel and accurately simulate the spatial and temporal characteristics of the read voltages.


Several models of voltage distributions supported by empirical measurements or simulated voltages in NAND flash have appeared in the literature. Cai et al.~\cite{bib::Cai2011MLCmodel} models the voltage distribution in 2-bit per cell MLC flash devices as a Gaussian distribution. Parnell et al.~\cite{bib:Parnell2014} proposed a parameterized Normal-Laplace mixture model that more accurately describes MLC flash read voltage distributions. Luo et al.~\cite{bib:LuoJSAC2016} proposed another accurate and computationally more efficient model for MLC flash, based on a modified version of the Student's t-distribution and a temporal power law. Liu et al.~\cite{bib::Liu2020VoltageGAN}  used a neural network to model simulated read voltages as a function of P/E cycles for one individual program level in isolated MLC flash cells. 
In addition, statistical analysis of hard bit error statistics in~\cite{bib::Taranalli2016MLCJournal, bib::Papandreou20193DTLC} and characterization of dominant error patterns~\cite{bib::Cai2012PatternModel} offer a valuable empirical understanding of flash memories. However, as effective as these models have been in the scenarios to which they have been applied, none has  provide an accurate model of the complex spatial and temporal characteristics of flash memory read voltages. 

Our goal in this paper is to use machine-learning techniques to develop a comprehensive and accurate model of the flash memory channel, with the aim of capturing both spatial ICI effects and temporal distortions arising from P/E cycling. 

Recently, generative modeling techniques~\cite{bib::Goodfellow2014GAN, bib::Kingma2014VAE} have been successfully applied to image processing~\cite{bib::Isola2017Pix2Pix}. In view of the demonstrated power of neural networks in learning complex multidimensional distributions, we propose the use of conditional generative nets as an approach to modeling flash memory read voltages in space and time. We adopt the conditional VAE-GAN as our training architecture, where the fusion of the variational autoencoder (VAE)~\cite{bib::Kingma2014VAE} and the generative adversarial network (GAN) ~\cite{bib::Goodfellow2014GAN} can leverage the information from the latent space to  produce high-quality, accurate reconstruction with the help of the  discriminator. 

We train the conditional VAE-GAN network  to regenerate the soft read voltage levels from program levels, using a dataset of measurements from a 1X-nm, 3-bit per cell TLC NAND flash memory. By incorporating the exact P/E cycle counts into the architecture, we are able to control the regenerated voltage levels at a specified time stamp.

We formulate a novel generative modeling workflow to generate soft read voltages from all program levels over a range of P/E cycles. When we train the network on measured data, this conditional method can regenerate realistic cell-level read voltages on an array of flash cells from a corresponding array of program levels and a given P/E cycle. To our knowledge, this is the first work that models both temporal features (P/E cycles) and spatial characteristics (ICI effects). We use two evaluation metrics to compare the model results to measured data: the read voltage distributions, i.e., the estimated probability density functions (PDFs), at different time stamps, and the relative frequencies of spatially-dependent and pattern-dependent ICI-induced hard read errors. Details of these experimental comparisons are provided in Section~\ref{sec::exp}.


\section{Flash Memory Basics}
\label{sec::basics}

\vspace{-0.5ex}
\subsection{Flash Structure and Experimental Procedure}
\label{subsec::structure}
	
	\begin{figure}
		\centering
			\includegraphics[width=1\columnwidth]{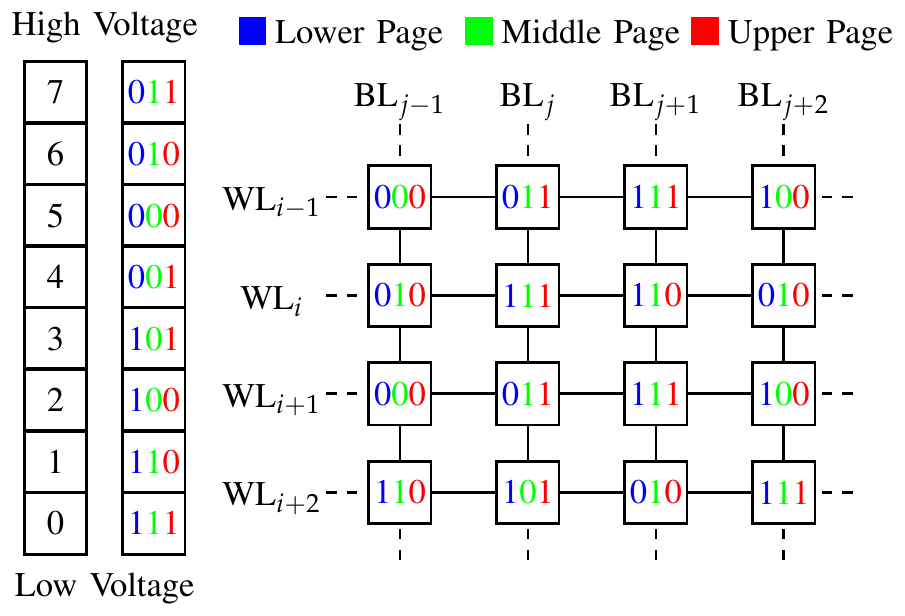}
		\vspace{-4ex}
		\caption{\small{(Left) Example mapping of cell program levels to binary representations in TLC flash. (Right) Schematic diagram of  a TLC flash memory block showing the 2-D array of cells connected in the horizontal direction by wordlines (WLs) and in the vertical direction by bitlines (BLs).}}
		\label{fig:flash_block}
		\vspace{-3ex}
	\end{figure}

The basic unit of data storage in NAND flash memory is a floating-gate transistor, referred to as a cell. Today's flash memories are capable of storing single or multiple bits (e.g. 2 to 5 bits) per cell, where the $n$-bit strings correspond to $2^n$ program levels. The cells are organized into an interconnected two-dimensional (2-D) array, called a block, via horizontal wordlines (WLs) and  vertical bitlines (BLs). The flash memory chip is composed of a collection of such blocks. Fig.~\ref{fig:flash_block} depicts a schematic diagram of a planar TLC flash memory block and an example of a mapping from program levels to binary bits. 

The basic unit of write (i.e., program) and read operations in flash memory is a page, corresponding to a logical bit position in a wordline of a block. On the other hand, the basic unit of an erase operation is an entire block. 


To characterize the channel and create datasets for machine learning, we conducted  program/erase (P/E) cycling experiments on several blocks of a commercial 1X-nm TLC flash memory chip. During each P/E cycle, the selected block is first erased, and then pseudo-random data are programmed into the successive pages in the block. At $4000$, $7000$, and $10000$ cycles, we perform read operation on the tested block. For each cell, we then the record program level and the measured voltage level. The P/E cycling experiments were performed at room temperature in a continuous manner with no wait time between the erase-program-read operations.

In the program operation, we refer to the program levels of three consecutive cells along WLs and BLs as a pattern. As an example, in Fig.~\ref{fig:flash_block}, the programmed levels $\text{PL}_{i-1,j}\text{PL}_{i,j}\text{PL}_{i+1,j}$ in WLs $(i-1),i,(i+1)$ of BL $j$, correspond to bit strings ``011'', ``111'', ``011'', which we associate with the pattern 707 in the vertical (BL) direction.

\vspace{-2ex}
\subsection{Spatio-temporal Characteristics}
\label{subsec::ST}

\begin{figure}[t]
	\begin{center}
		\includegraphics[width=0.80\columnwidth, trim={0in 1in 0in 0in}]{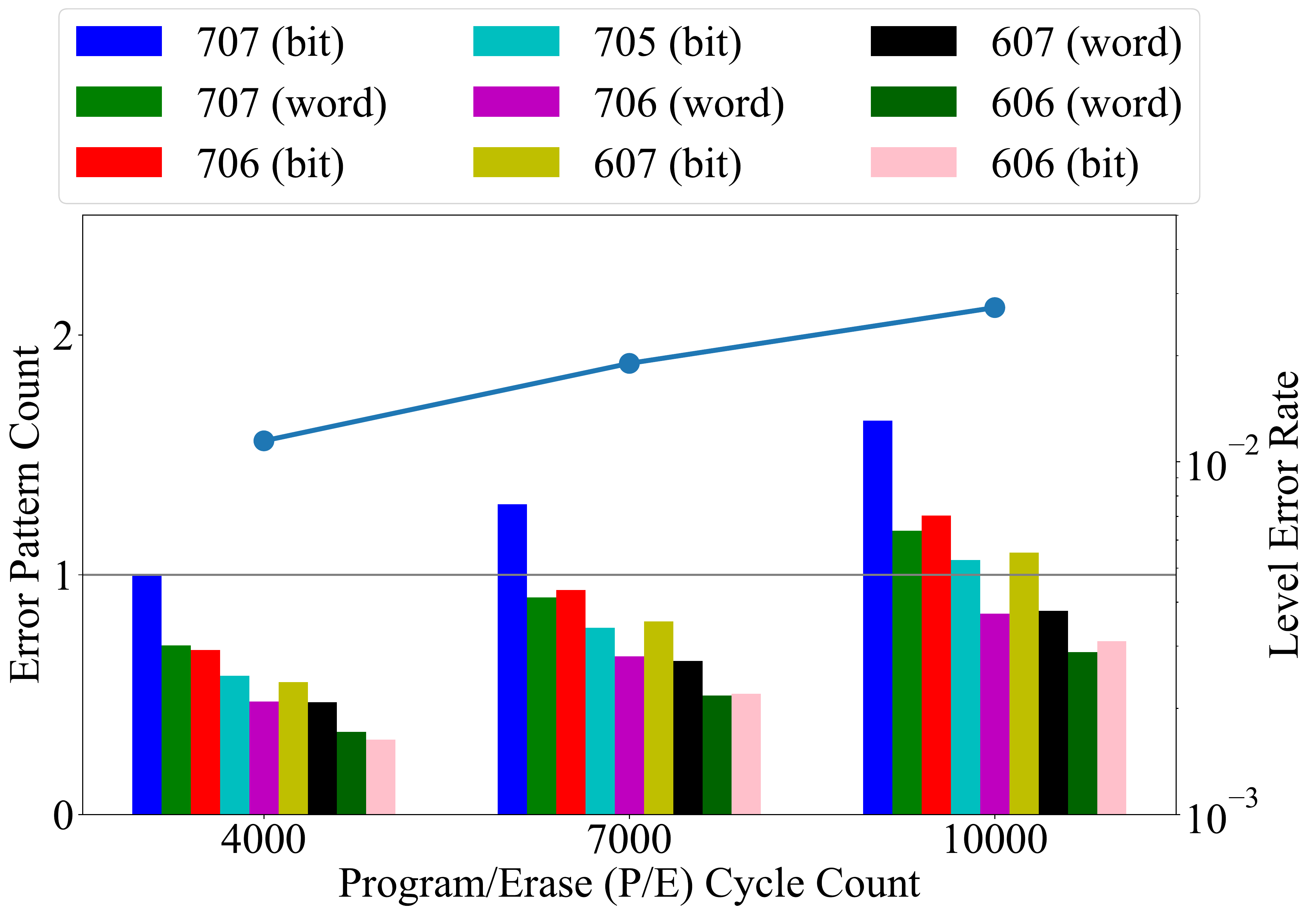}
	\end{center}
	\caption{Count of top error-prone patterns and level error rate at selected P/E cycle. The error pattern counts are normalized by the count of pattern 707 at bitline direction of 4000 P/E cycle.}
	\label{fig::ler_pattern}
	\vspace{-2ex}
\end{figure}

Repeated P/E operations induce wear on the flash cells,  and high-low-high program  levels in three consecutive cells often leads to severe ICI effects. These distortions represent the spatio-temporal nature of the flash memory channel. As a result, the  level error rate increases as the temporal P/E cycle count grows, as seen in Fig.~\ref{fig::ler_pattern}. 

The spatial ICI effect refers to the phenomenon where programming of a cell induces changes in the voltage levels of neighboring cells within its block. In particular, the read voltage level  of a cell programmed to a low level may be inadvertently increased if its adjacent cells are programmed to high levels, i.e., when the programming pattern is high-low-high. For example, if we program a 707 pattern in a TLC flash memory, the  read voltage of the low central cell may be increased by its high adjacent cells. During data detection, the recovered program level of the central ``victim'' cell may therefore be erroneously interpreted as an incorrect level. These the severe ICI effects can be observed in Fig.~\ref{fig::ler_pattern}. At each P/E cycle, we see that the cell errors are not randomly distributed; they are clearly affected by  neighboring program levels. The 9 patterns shown are the most error-prone in both WL and BL directions. Pattern 707 in the BL direction is the most severely affected by ICI. Moreover, patterns 707, 706, and 607 in the BL direction are more error-prone than those on the WL direction. 

We note that, to mitigate the effects of ICI in flash memory, the use of constrained codes that prevent the appearance of ICI-prone patterns has been proposed; see, for example,~\cite{bib::Qin2014ConstrainICI, bib::Ahmed2021Flashcode, bib::Ahmed2022RRcode}. Accurate modleing  of the dependence of WL and BL pattern errors on the  P/E cycle count can be  a valuable tool to help researchers design efficient, time-aware constrained codes. 

\section{Generative Modeling for Flash Memory}
\label{sec::GM}

\begin{figure}[t]
	\begin{center}
		\includegraphics[width=1\columnwidth, trim={0in 0.6in 0in 0.1in}]{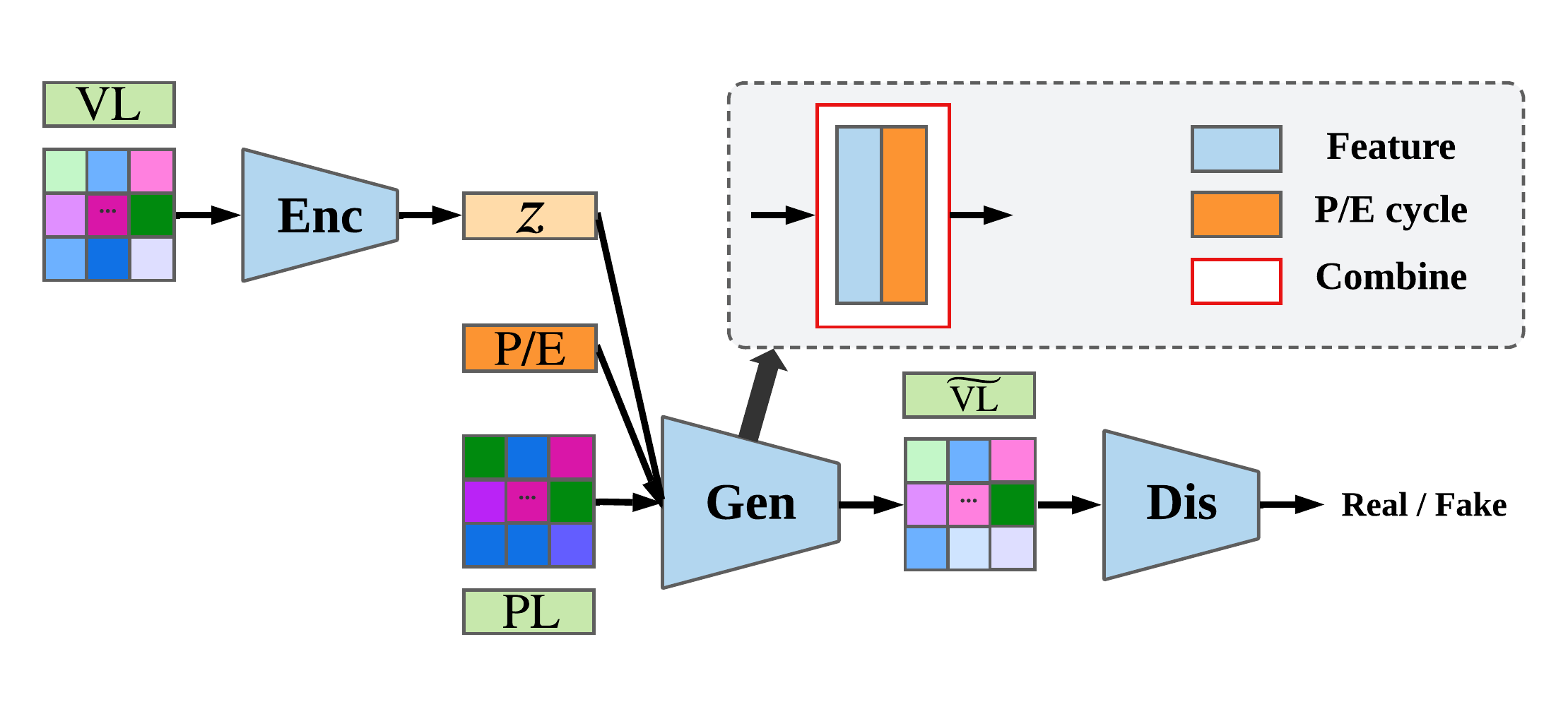}
	\end{center}
	\caption{Generative modeling pipeline: encoder, generator, and discriminator constitute the generative modeling workflow. Here, $z$ is the latent vector; $\text{P/E}$ is the corresponding P/E cycle count; $\text{PL}$ is the array of program levels, $\text{VL}$ is the array of measured read voltage levels, and $\widetilde{\text{VL}}$ is the reconstructed array of read voltage levels. In our implementation, $\text{PL}$, $\text{VL}$, and $\widetilde{\text{VL}}$ are all $64\times 64$ arrays.}
	\label{fig::cvaegan}
	\vspace{-2ex}
\end{figure}

In this section, we discuss our generative modeling pipeline for the flash memory channel. We adopt a conditional VAE-GAN (cVAE-GAN) architecture~\cite{bib::Larsen2016VAEGAN} for our pipeline, depicted in Fig.~\ref{fig::cvaegan}. Our goal is to learn a mapping between program levels and soft read voltage levels at various P/E cycles, where the reconstructed voltage levels accurately reflect the spatial and temporal nature of the  channel. 




\vspace{-2ex}
\subsection{Generative Flash Modeling}
\label{subsec::GMsub}

Using the P/E cycling experiment outlined in Section~\ref{subsec::structure}, we collect the paired channel instances at specific P/E cycles, where the channel instances are denoted as $\{(\text{PL}, \text{VL}, \text{P/E})\}$. Here, we use  $\text{PL}$, $\text{VL}$, and $\text{P/E}$ to denote the input program level of channel, the output voltage level of channel, and the P/E cycle count, respectively. The goal of channel reconstruction and generative modeling in this paper is to learn the analytically intractable likelihood $P(\text{VL}|\text{PL},\text{P/E})$, with the aim of capturing the spatio-temporal nature of the flash memory channel. 

Fig.~\ref{fig::cvaegan} summarizes the architecture of the generative modeling pipeline. The conditional VAE-GAN architecture consists of three components: an encoder ($Enc$), a generator ($Gen$), and a discriminator ($Dis$). The encoder maps the read voltages to the latent vector $z$ at a specific $\text{P/E}$ and replaces the prior distribution $P(z)$ in the GAN with the learned posterior distribution $P(z|\text{VL}, \text{P/E})$. The decoder in the VAE shares its weights with the GAN generator~\cite{bib::Goodfellow2014GAN}. In the conditional setting, the variational lower bound of $P(\text{VL}|\text{P/E})$ can be derived as

\vspace{-2ex}
{
\begin{align*}
\log P(\text{VL}|\text{P/E})\geq & -D_{KL}(Q(z|\text{VL},\text{P/E})||P(z|\text{P/E})) \\
&+\mathbb{E}_{Q(z|\text{VL},\text{P/E})}[\log (P(\text{VL}|z,\text{P/E}))] \\
\label{eq::vae}
\end{align*}
}

\vspace{-4ex}
\noindent
where $D_{KL}$ represents the Kullback-Leibler (KL) divergence. The distribution $Q(z|\text{VL},\text{P/E})$ of the latent vector $z$ is trained to approach $P(z|\text{P/E})$ via the KL loss $\mathcal{L}_{KL}$, where $P(z|\text{P/E})$ is assumed to be a Gaussian distribution.



Generator $Gen$ will take both the learned latent vector and $\text{PL}$ as input and generate a ``fake'' $\widetilde{\text{VL}}$. The latent vectors are sampled from $Q(z|\text{VL},\text{P/E})$ using the re-parameterization trick~\cite{bib::Kingma2014VAE}. When sampling different latent vectors $z$ from the same distribution, we can generate multiple arrays of plausible voltages levels. The variations in these output arrays for a given array of program levels reflect the stochasticity of the channel. The discriminator measures the difference between $\text{PL}$ and $\widetilde{\text{VL}}$. The loss in the conditional GAN part is
\begin{equation*}
\small
\begin{aligned}
\mathcal{L}_{GAN} = \log(1-Dis(\text{PL}, Gen(\text{PL}, \text{P/E}, z))) + \log(Dis(\text{PL}, \text{VL})).
    \label{eq::gan}
\end{aligned}
\end{equation*}
\noindent 
Similar to VAE-GAN~\cite{bib::Larsen2016VAEGAN}, we encourage the reconstructed voltage levels to match the authentic voltage levels, using the $\ell_2$-norm to measure the reconstruction loss 
\vspace{-1ex}
\begin{equation*}
    \mathcal{L}_{recon} = ||\text{VL}-Gen(\text{PL},\text{P/E}, z)||_2.
    \label{eq::l1}
    \vspace{-1ex}
\end{equation*}
\noindent 
Combining these equations, we formulate the loss function of the cVAE-GAN architecture as
\begin{equation}
    \min_{Gen,En}\max_{Dis}\mathcal{L}_{GAN}+\alpha\mathcal{L}_{recon}+\beta\mathcal{L}_{KL}.
    \label{eq::allloss}
    \vspace{-1ex}
\end{equation}

\vspace{-2ex}
\subsection{Spatio-temporal Combination}
\label{subsec::stcombine}

Aiming to capture the spatial ICI effects in the channel model, we implement the generator using a convolutional neural network (CNN) in $Gen$, where $\text{VL}$ is reconstructed from the $\text{PL}$ values in its cell and neighboring cells. To generate $\text{VL}$ at an explicit P/E cycle count, we control the generator with an additional temporal factor and incorporate the explicit P/E cycle count into $Gen$.

We first encode the normalized P/E cycle count into a $d$-dimensional P/E vector, which contains expressive powers of the normalized P/E cycle, e.g., $\text{P/E}^{2}$, $\sqrt{\text{P/E}}$, etc. Then, we spatially replicate the $d$-dimensional P/E vector to the feature map with appropriate size $H\times W\times d$ and concatenate it with the $H\times W\times C$ feature from each layer in $Gen$, where $H\times W$ is the spatial dimension of the feature from each convolutional layer and $C$ is the number of channels in the CNN. The channel-wise combination produces the final feature with size $H\times W\times (C+d)$ of each layer. The fusion of the features from the program levels and the P/E feature maps guarantees the spatial-temporal characteristics of the reconstructed voltage levels.


\vspace{-2ex}
\subsection{Implementation Details}
\label{subsec::implemention}

We implement and evaluate our method using the recorded data in a commercial TLC flash chip. With the generative modeling framework, we believe this data-driven approach can be  flexibly applied  to flash memories of any technology generation and scale. In TLC flash memory, each block contains hundreds of pages, each of which is typically 8-16 kB in size. In order to represent the TLC flash memory without bias, we collect data from several blocks of one 1X-nm TLC chip at selected P/E cycle counts. We crop the blocks into non-overlapping $64\times 64$ 2-D arrays to formulate our paired data.  The number of 2-D arrays in the training set is $1.5\times 10^5$ ($5\times 10^4$ for each P/E cycle) and the size of the evaluation dataset is $2.1\times 10^4$ ($7\times 10^3$ for each P/E cycle). The dimensions of latent vector $z$ and P/E cycle vector are both set to 6.
\begin{remark}
\label{remark::network}
We adapt the conditional VAE-GAN in \cite{bib::Zhu2017BiCycleGAN} for input arrays of $64\times 64$ program levels and output arrays of $64\times 64$ voltage levels, each of which has a single channel. Three network modules in Fig.~\ref{fig::cvaegan} refer to: ResNet \cite{bib::He2016ResNet} ($Enc$), U-net \cite{bib::Ron2015Unet} ($Gen$), and PatchGAN \cite{bib::Isola2017Pix2Pix} ($Dis$). The following descriptions of the modules exploit the terminologies in the corresponding references. 
\begin{enumerate}
    \item Encoder: We use the two residual blocks, each of which contains two $3\times 3$ convolutional layers with stride 1 and padding 1. We then add two linear layers, which map output features to mean and variance for the latent vector.
    \item Generator: $Ck$ denotes a Convolution-BatchNorm-ReLU layer with $k$ output channels. All convolutions are $4\times 4$ kernels applied with stride 2 and padding 1. The network architecture before spatio-temporal combination can be described as 
    $$
    \begin{aligned}
    &(\text{Down Part})\ C64,C128,C256,C512,C512,C512 \\
    &(\text{Up Part})\ C512,C512,C256,C128,C64,C1
    \end{aligned}
    $$
    where we inject latent vector $z$ by replication and concatenation into every layer in the ``Down'' part \cite{bib::Zhu2017BiCycleGAN}, and each layer in the ``Up'' part receives skip connections from the corresponding  layer  in the ``Down'' part \cite{bib::Ron2015Unet}.
    \item Discriminator $Dis$: The input to the discriminator is the concatenation of fake voltage levels and program levels. With the same naming convention as in the generator, we express the discriminator as $C64, C128, C1$.
\end{enumerate}
\end{remark}

\vspace{-3ex}
\begin{remark}
The training details of the generative modeling methods are as follows. Adam optimizer is used with learning rate $2\times 10^{-4}$. Parameters in the loss function (\ref{eq::allloss}) are set to $\alpha=10$ and $\beta=0.01$. We train the conditional VAE-GAN for 7 epochs with batch size 2.

During evaluation, we use program levels and latent vector $z$ sampled from a standard Gaussian distribution. For each program level array, we prepare $10$ different sampled latent vectors to evaluate the learned model.
\end{remark}

\section{Experimental Results}
\label{sec::exp}

\begin{figure*}[tp]
	\centering
	\begin{subfigure}[b]{0.90\textwidth}
		\includegraphics[width=\linewidth, trim={0in 8in 0in 0in}]{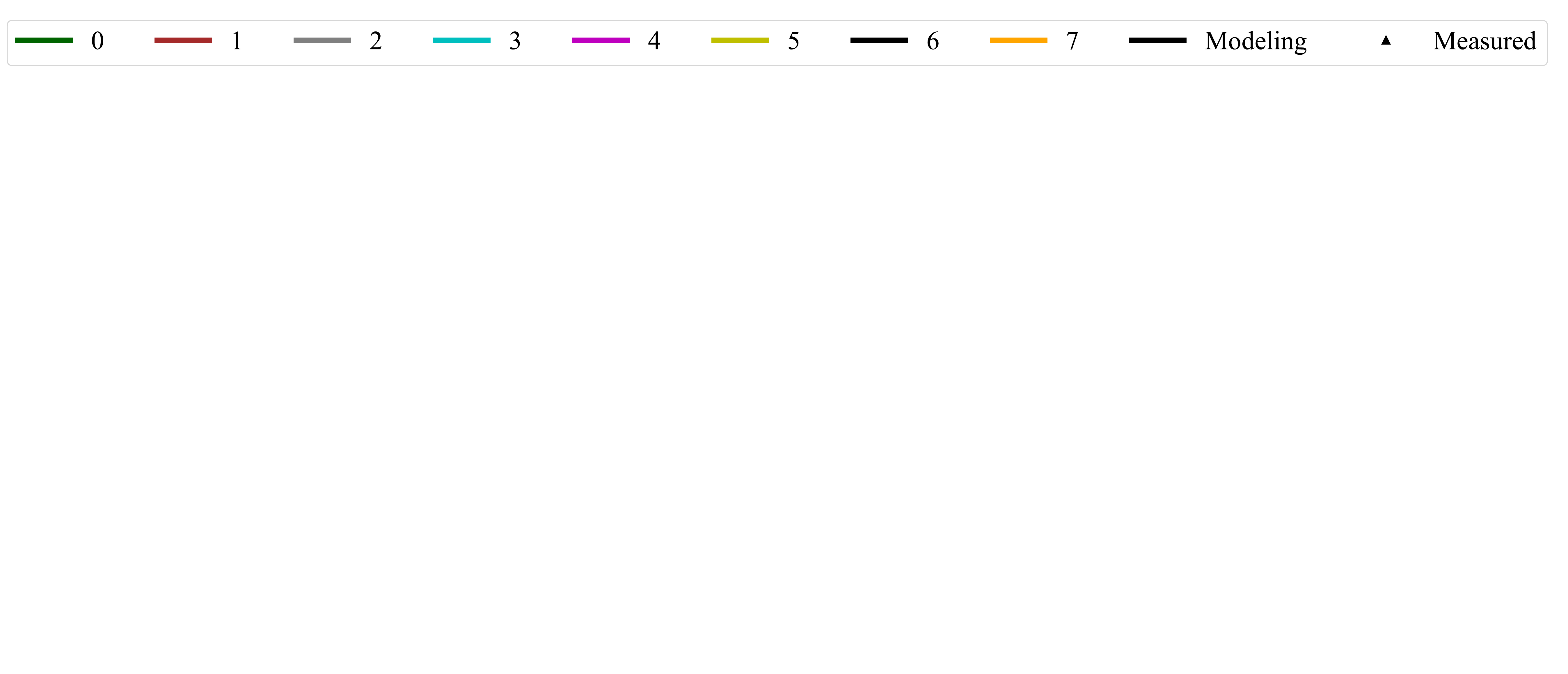}
	\end{subfigure}
	\begin{subfigure}[b]{0.30\textwidth}
		\includegraphics[width=\linewidth]{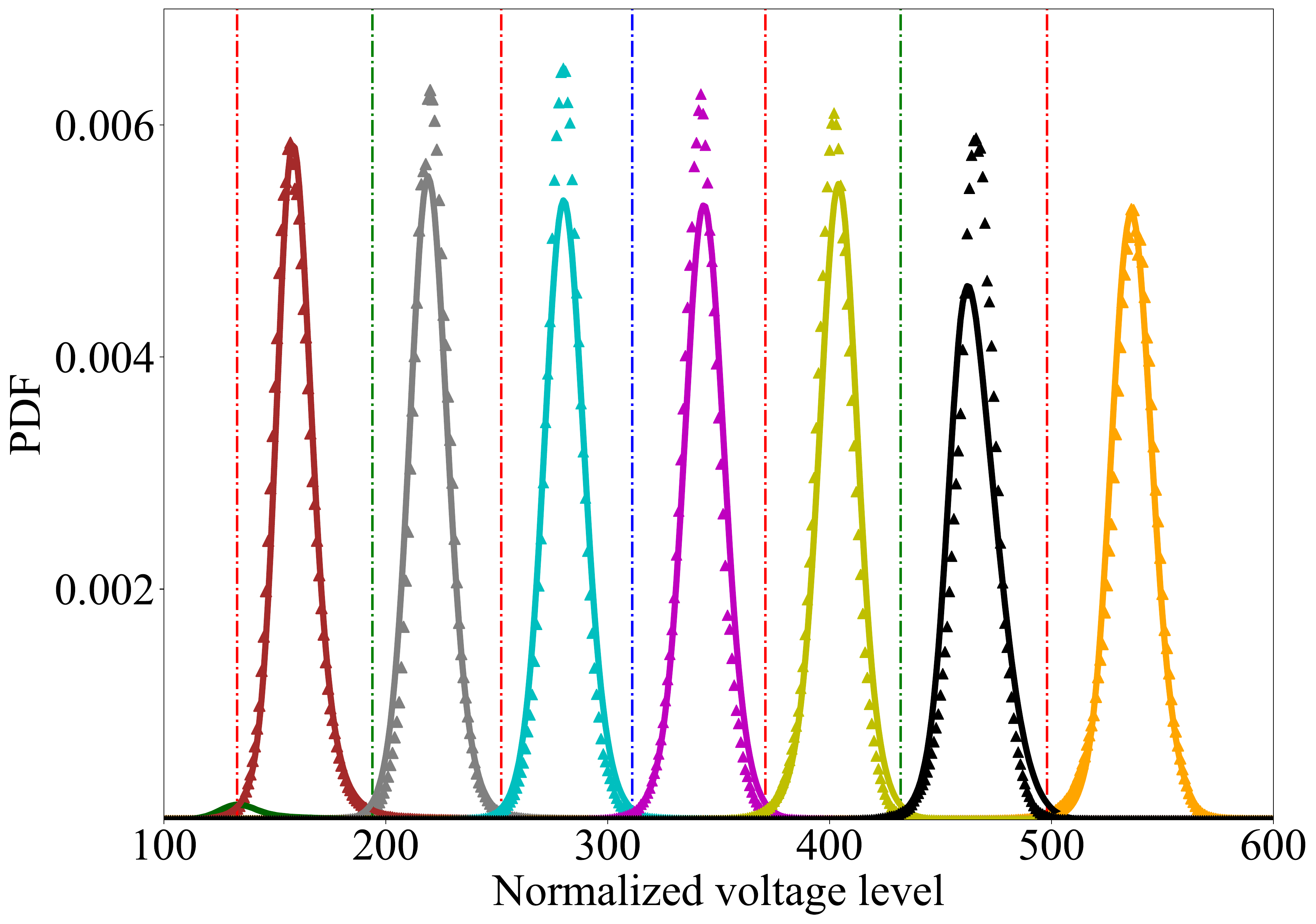}
		\caption{4000 P/E cycles}
	\end{subfigure}
	\begin{subfigure}[b]{0.30\textwidth}
		\includegraphics[width=\linewidth]{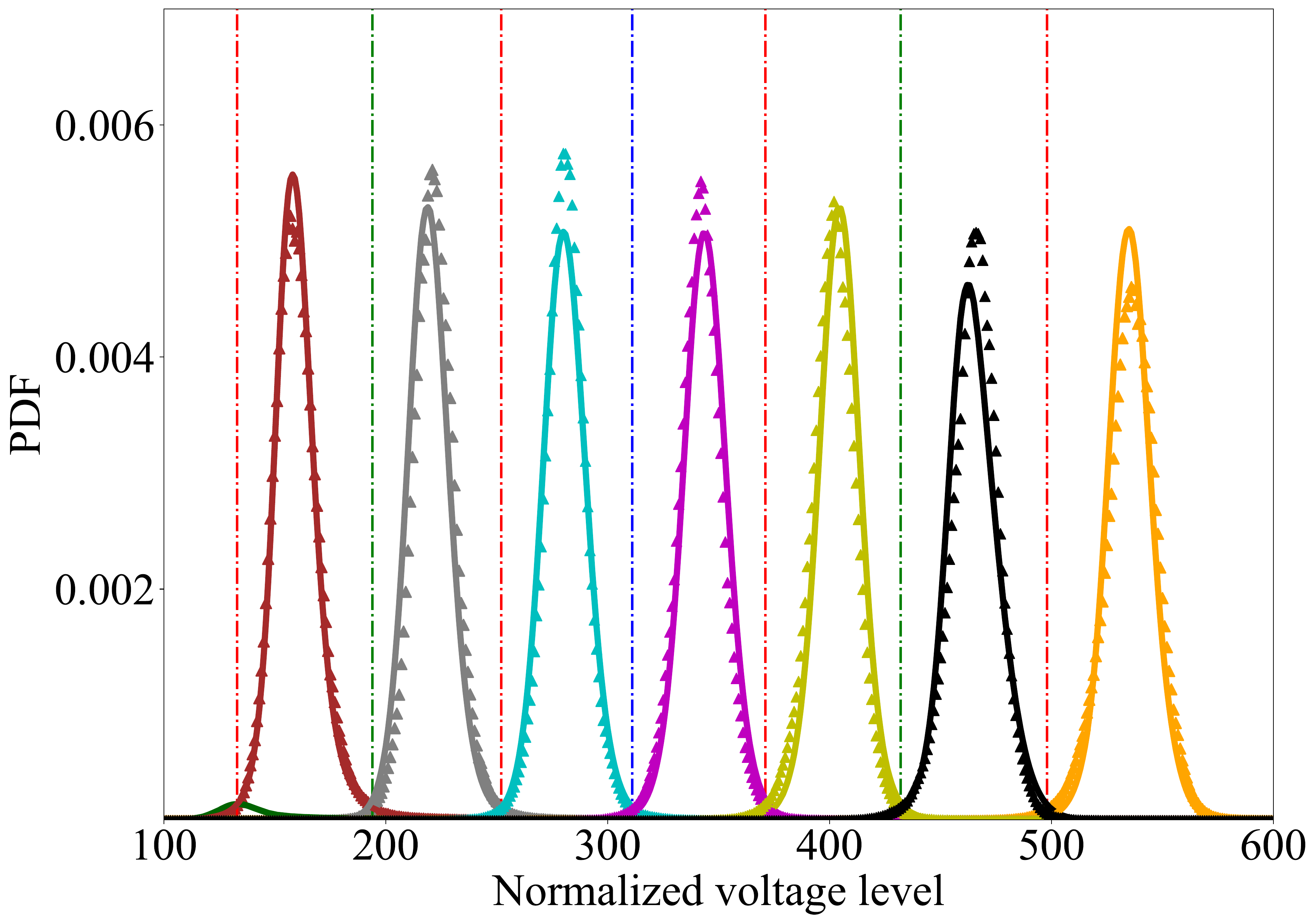}
		\caption{7000 P/E cycles}
	\end{subfigure}
	\begin{subfigure}[b]{0.30\textwidth}
		\includegraphics[width=\linewidth]{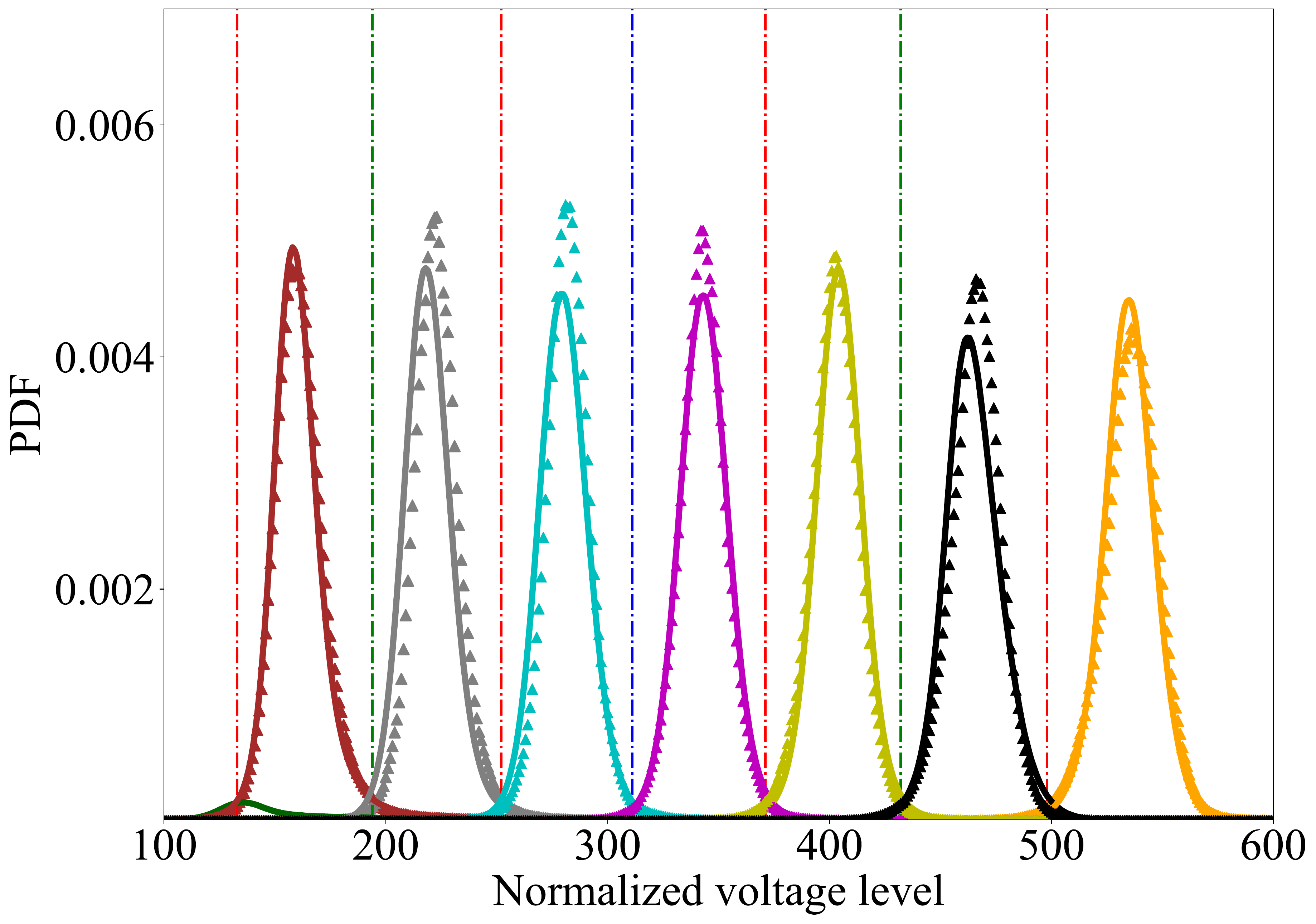}
		\caption{10000 P/E cycles}
	\end{subfigure}
	\vspace{1ex}
	\caption{PDF visualizations for measured and cVAE-GAN voltage levels at 4000, 7000, 10000 P/E cycles. In each sub-figure, solid curves represent cVAE-GAN modeled distribution and triangles represent measured distribution. The plots are in linear scale. Vertical dash-dotted lines are  fixed default voltage thresholds.}
	\label{fig::allcomppdf}
	\vspace{-3ex}
\end{figure*}

\begin{figure}[t]
	\begin{center}
		\includegraphics[width=1\columnwidth, trim={0in 1in 0in 0.10in}]{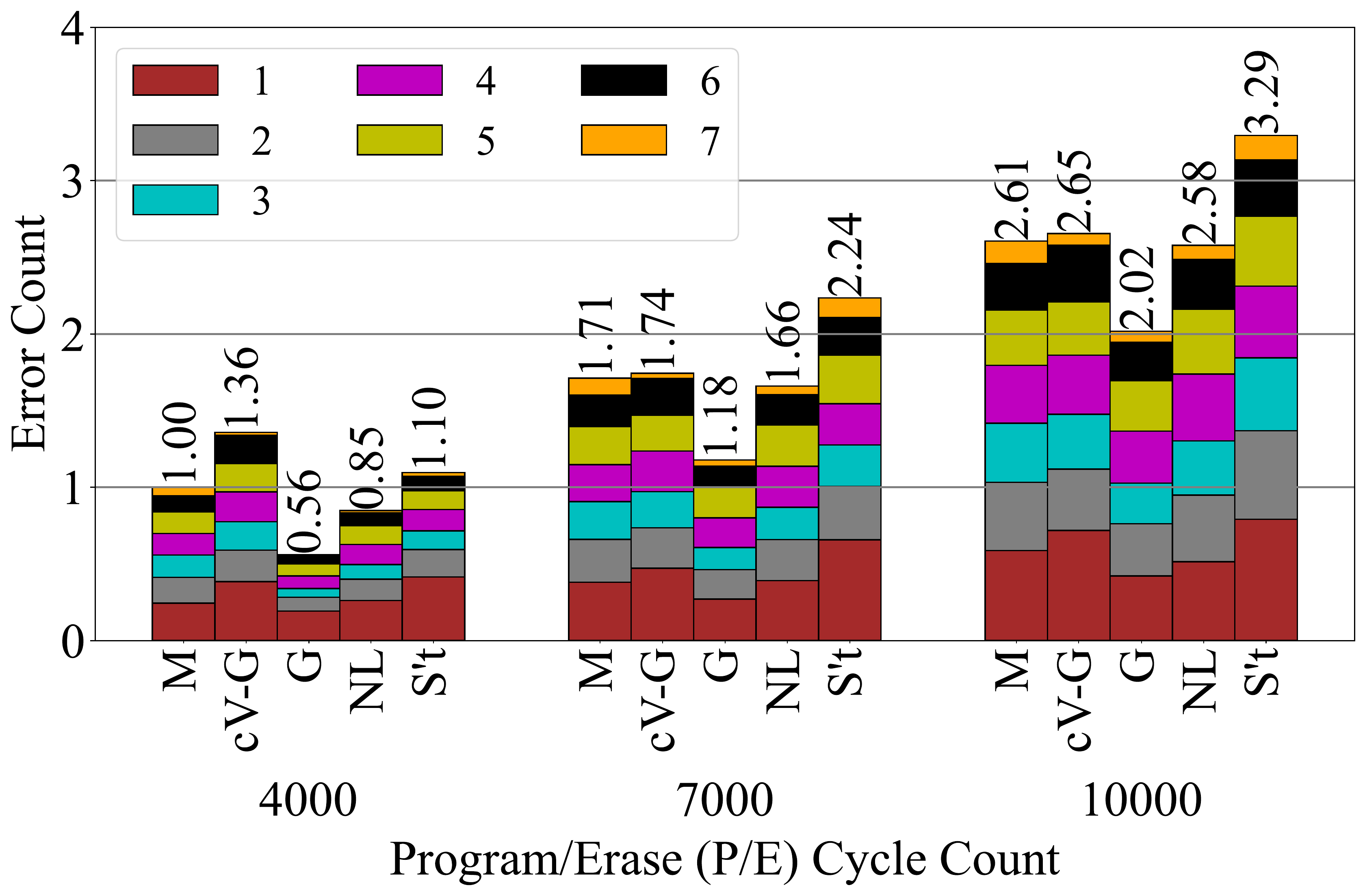}
	\end{center}
	\caption{Total error counts of measured (`M'), cVAE-GAN ('cV-G'), Gaussian (`G'), Normal-Laplace (`NL'), Student's t (`S't') model.In each bar, we stack the errors from program level 1 to program level 7. We normalize the error counts of measured data at 4000 P/E cycle as 1.}
	\label{fig::level_error_rate}
	\vspace{-4ex}
\end{figure}

To evaluate the quality of the reconstructed voltage levels and analyze the spatio-temporal nature of flash memory channel, we summarize our evaluation approaches as follows. 
\begin{enumerate}
\item Distribution: The frequency of occurrence of each voltage level given the program level and P/E cycle count is used to estimate the conditional probability of that level and time. We visualize the PDFs for measured data and reconstructed data. For quantitative comparison, we compare the distributions generated by our data-driven method with three statistical fitting methods~\cite{bib:LuoJSAC2016}, using the metric of level error count.
\item Inter-cell interference (ICI): For cells programmed to 0 level that suffer an error according to their voltage level, we compute the relative frequencies of the patterns of program levels in adjacent cells in the WL and BL directions. We visualize these relative error frequencies using pie charts. 
\end{enumerate}

\begin{remark}
We compared the cVAE-GAN model to other popular generative modeling architectures: conditional GAN \cite{bib::Isola2017Pix2Pix}, conditional VAE \cite{bib::Sohn2015CVAE}, and Bicycle GAN \cite{bib::Zhu2017BiCycleGAN}. The networks were adapted for the array sizes of program levels and voltage levels,  as in Remark~\ref{remark::network}. We then compare the learned distributions with the measured distribution using the total variation distance $d_{TV}$,
$$
\vspace{-0.5ex}
    d_{TV}(P_{real}, P_{fake}) = \frac{1}{2}\sum_{\text{VL}}|P_{real}(\text{VL})-P_{fake}(\text{VL})|
\vspace{-0.5ex}
$$
The numerical results of $d_{TV}$ indicated that the cVAE-GAN achieved the smallest total variation distance with respect to the measured distributions. Therefore, we selected cVAE-GAN as our model for the present study. 
\end{remark}
\vspace{-1ex}

\vspace{-2ex}
\subsection{Distribution Analysis}
\label{subsec::pdf}

As we evaluate our learned model using input arrays of program levels, we collect regenerated voltage levels and count the frequency of occurrence of voltage levels over the voltage range. We then estimate the conditional PDFs of voltages associated with each program level and given P/E cycle. 

Fig.~\ref{fig::allcomppdf} shows the conditional PDFs for measured data and regenerated data in the evaluation dataset at three different time stamps. The $x$-axis represents the soft read voltages spanning a certain voltage level range. The $y$-axis represents the conditional PDF. We only show conditional PDFs from program level 1 to 7 due to normalization problems of program 0;  the detailed analysis of level 0 will be discussed in Section \ref{subsec::iciexp}. We make two observations from Fig.~\ref{fig::allcomppdf}. First, as P/E cycle count increases, the peak of the distribution in each program level drops and the distribution becomes wider. This indicates that more voltage levels exceed the read thresholds and more errors will be detected, which is consistent with the increased error rate in Fig.~\ref{fig::ler_pattern}. Second, our modeled data (solid curves) almost match the measured data (triangle markers) and, thus, capture the dependence on P/E cycles. 

For a more quantitative assessment, we compare our generative model with three state-of-the-art statistical models: Gaussian model~\cite{bib::Cai2011MLCmodel}, Normal-Laplace model~\cite{bib:Parnell2014}, and Student's t-distribution model~\cite{bib:LuoJSAC2016}. Following the optimization process used in \cite{bib:LuoJSAC2016} for an MLC flash device, we fit those statistical distributions to our TLC measured distributions. We minimize the KL divergence between real distribution and fake distribution by using the Nelder-Mead simplex method ~\cite{bib::Nelder1965Optimal}, where the KL divergence is denoted as $D_{KL}(P_{real},P_{fake})$. We obtain the best-fit parameters for all program levels, except $\text{PL}=0$, with each of the statistical distributions.

We then compute the level error counts under each of the distributions and quantitatively compare those fake distributions with real distributions, where the error count is a measure of the reliability and endurance of the flash memory. To calculate the level error counts from distributions, we fix 7 default read thresholds, as shown by the dash-dotted vertical lines in Fig. \ref{fig::allcomppdf}. The hard read voltages are determined by comparing soft voltages to these thresholds. For instance, if a voltage level of program level 1 lies below the first threshold or above the second threshold, the hard read level of the cell will not be designated as 1. The error probability of $\text{PL}=1$ is denoted as
$$
P(VL<V_{th(01)}|\text{PL}=1)+P(VL>V_{th(12)}|\text{PL}=1)
$$
where $V_{th(01)}$ denotes the voltage threshold used to distinguish between level 0 and level 1, and $V_{th(12)}$ denotes the voltage threshold used to distinguish between level 1 and level 2.

We present the level error counts for five models in the form of bar charts Fig.~\ref{fig::level_error_rate}. The $x$-axis represents the chosen P/E cycle count and the label directly under each bar represents the corresponding model name. The $y$-axis corresponds to the normalized error count. At each P/E cycle count, for each model,  the errors from 7 program levels are stacked as one individual bar. The stacked bar represent the total error count. For the measured distributions, we see that level 1 has the highest error counts and the total error count at 10000 P/E cycles is around 2.5$\times$ that at 4000 P/E cycles. 

For the statistical distributions, we observe that the Gaussian model does not estimate the error well; this is because the tails in the actual distribution  are becoming heavier as the device is cycled to severe conditions. The Normal-Laplace model, on the other hand, takes the heavier tails into consideration and provides accurate estimations of error counts at each P/E cycle. Student's t-distribution over-estimates the errors at those P/E cycles. For the machine learning approach, cVAE-GAN slightly overestimates the wear conditions in 7000 and 10000 P/E cycles. At 4000 P/E cycles, cVAE-GAN produces more errors than the measured data. In conclusion, Normal-Laplace is the best statistical model to capture the distributions of flash devices. Our cVAE-GAN works better than Normal-Laplace at 7000 and 10000 P/E cycles but overestimate the errors at 4000 P/E cycles. (However, as shown in the next section, the generative model can not only generate realistic-looking voltage distributions but also accurately learn spatial characteristics of the read voltages.)

Overall, we conclude that cVAE-GAN can regenerate voltages as a function of P/E cycles and produce high-quality voltage levels according to visual and quantitative metrics. 

\vspace{-2ex}
\subsection{Characterization of ICI Effects}
\label{subsec::iciexp}

As discussed in Section~\ref{subsec::ST}, the read voltage of a cell may be adversely increased by ICI when the neighboring cells are programmed to high levels. Cells programmed to level 0 are the most susceptible to such ICI effects. Generating voltage levels with ICI effects is complicated due to the pattern-dependent and spatially-dependent distortions. We remark that classical statistical models focus on regeneration of the PDFs of the measured data and, as such, are not expected to be effective in capturing ICI effects. 

We evaluate how well the generative model learns spatial ICI properties by examining errors associated with program level  patterns
$\text{PL}_{i,j-1} \; \text{PL}_{i,j}  \; \text{PL}_{i,j+1}$ and $\text{PL}_{i-1,j} \;\text{PL}_{i,j}   \; \text{PL}_{i+1,j}$ in the WL and BL directions, respectively. As we observed in Fig.~\ref{fig::ler_pattern}, the most error-prone patterns have central victim cell as 0, i.e., $\text{PL}_{i,j}=0$. We consider pattern-dependent error probabilities in both directions. The error probability measures the relative frequency of occurrence of the WL and BL patterns when an error occurs in the victim cell. More precisely, we calculate the pattern-dependent error probabilities in WL and BL directions,
$$
\begin{aligned}
P(\text{PL}_{i,j-1},\text{PL}_{i,j}=0,\text{PL}_{i,j+1}|\text{VL}_{i,j}>V_{th(01)},\text{PL}_{i,j}=0) \\
P(\text{PL}_{i-1,j},\text{PL}_{i,j}=0,\text{PL}_{i+1,j}|\text{VL}_{i,j}>V_{th(01)},\text{PL}_{i,j}=0).
\end{aligned}
$$

\begin{figure}[t]
\centering
	\begin{subfigure}[b]{1\columnwidth}
		\includegraphics[width=\linewidth, trim={0in 5.2in 0in 0.10in} ]{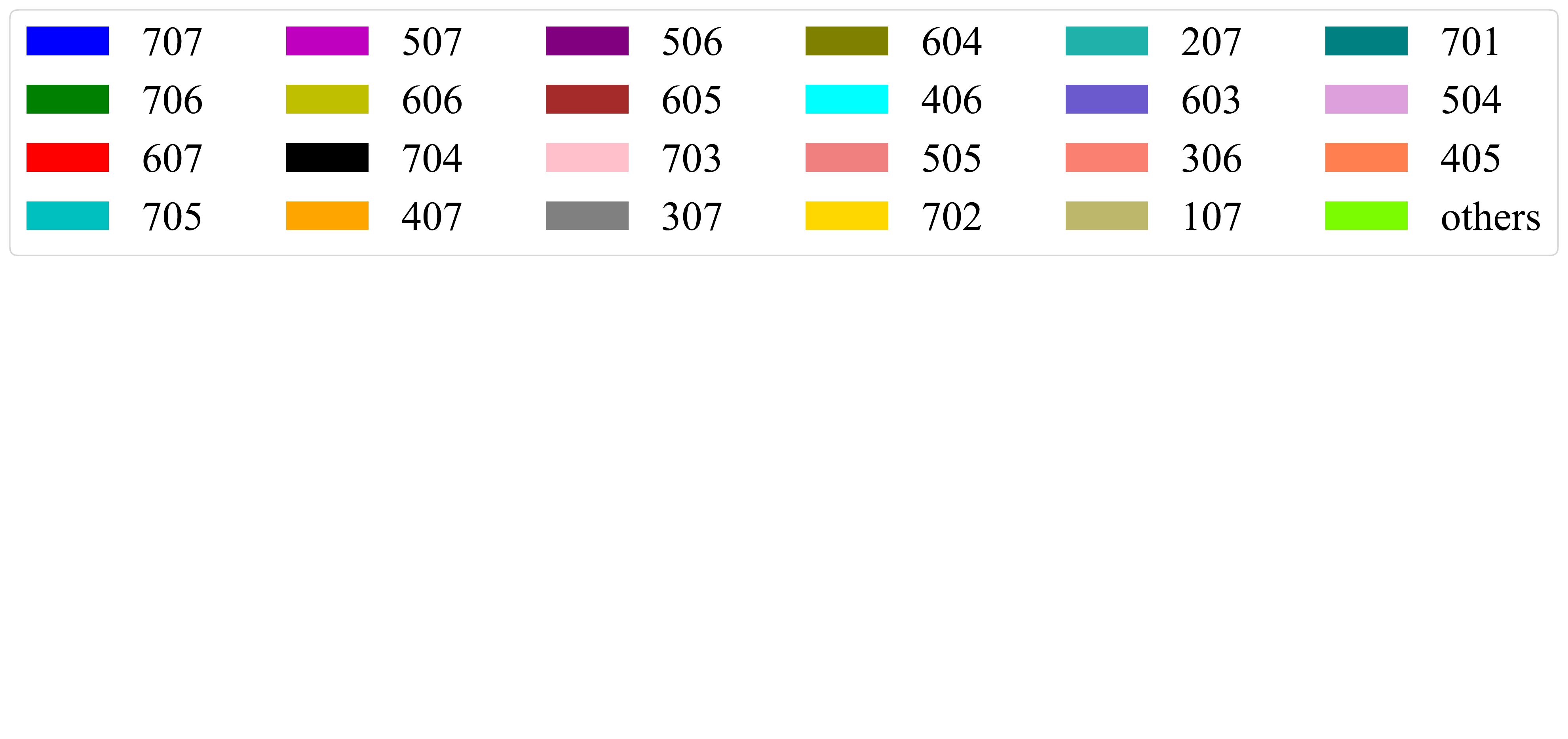}
	\end{subfigure}
	\begin{subfigure}[b]{1\columnwidth}
	    \centering
	    \captionsetup{justification=centering}
		\includegraphics[trim={0in 0.7in 0in 0.0in}, width=0.38\linewidth]{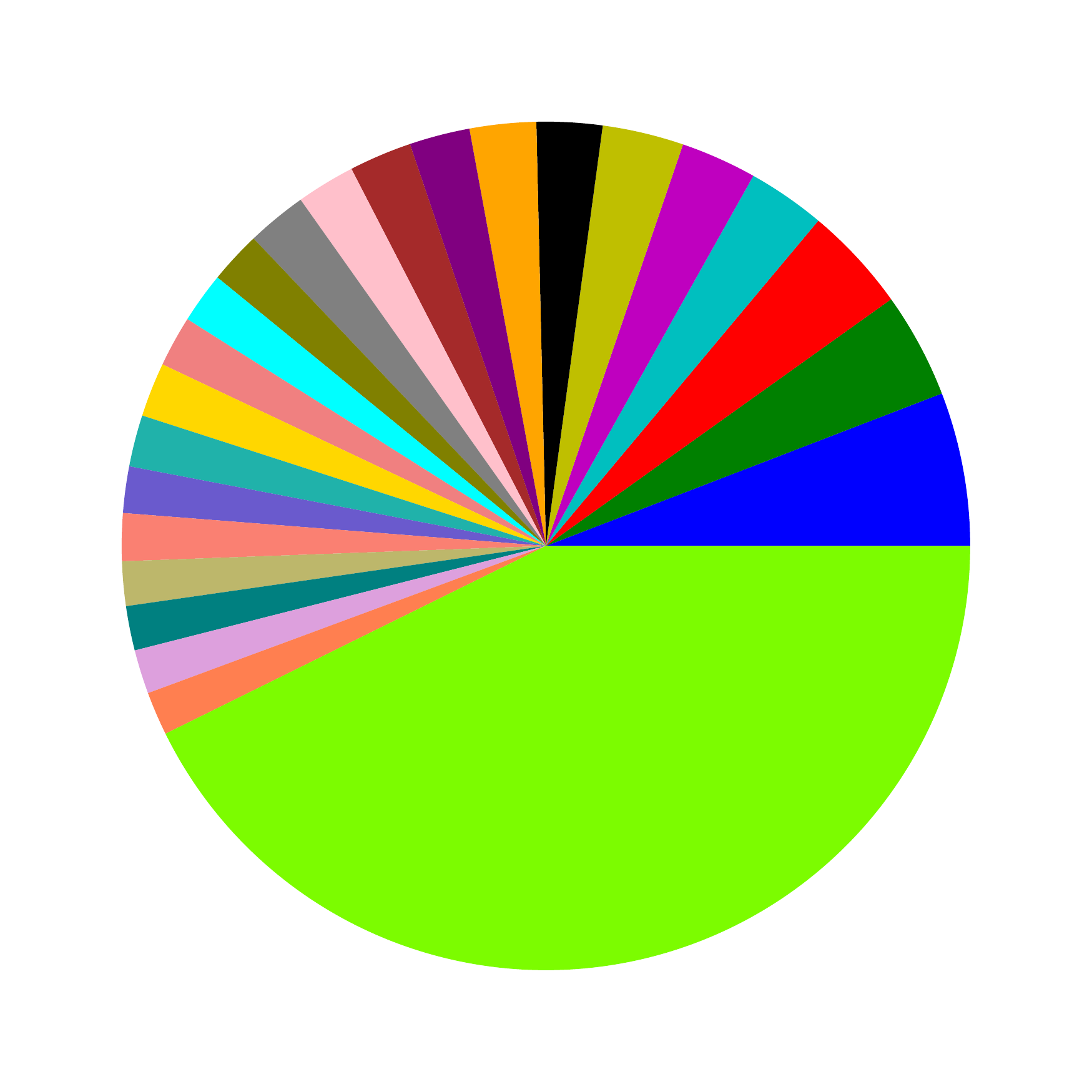}
		\hspace{0.05\columnwidth}
		\includegraphics[trim={0in 0.7in 0in 0.0in}, width=0.38\linewidth]{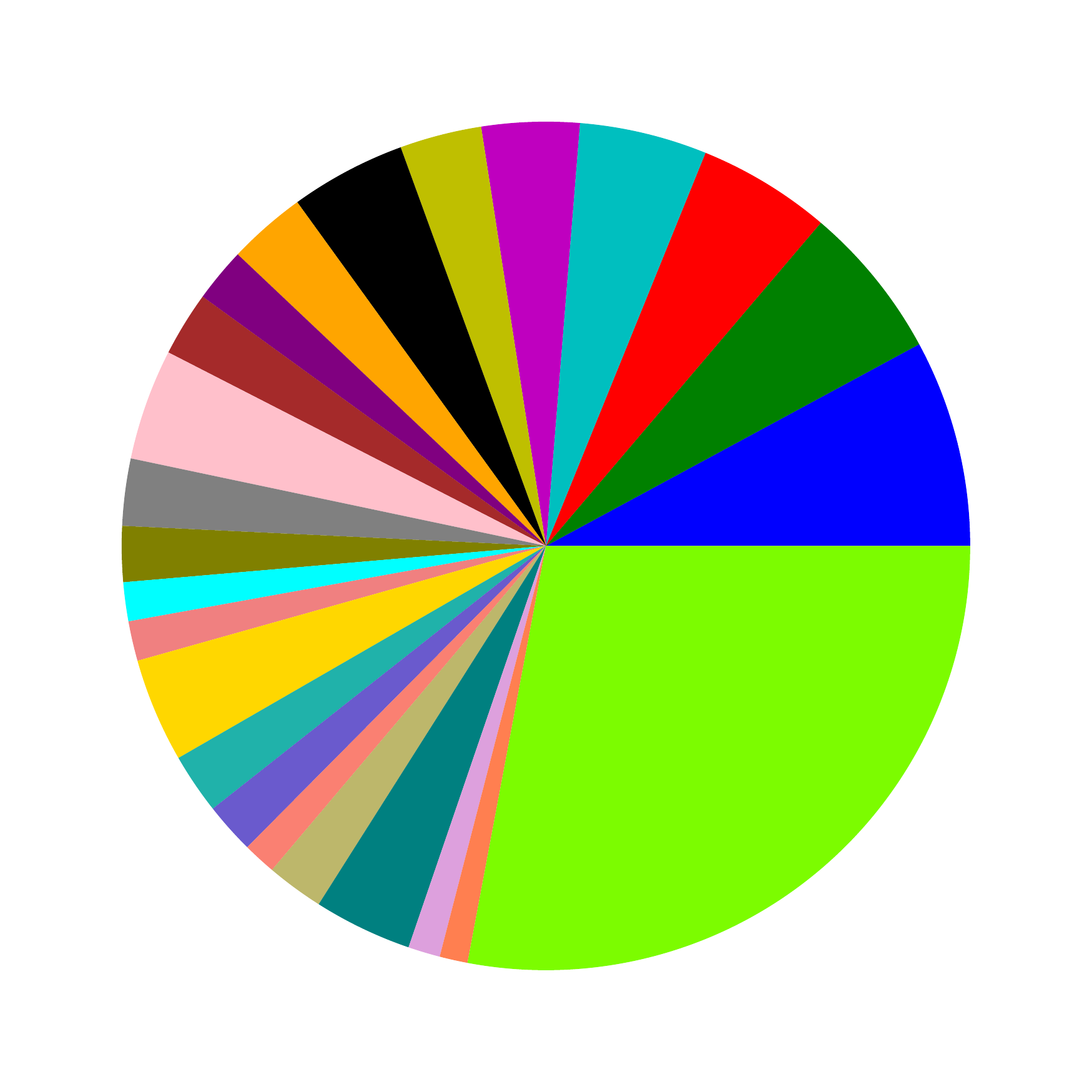}
		\caption{Measured errors in WL (left) and BL (right) \\ 97921 total errors observed at $\text{PL}=0$}
	\end{subfigure}
	\begin{subfigure}[b]{1\columnwidth}
	    \centering
	    \captionsetup{justification=centering}
		\includegraphics[trim={0in 0.7in 0in 0.0in}, width=0.38\linewidth]{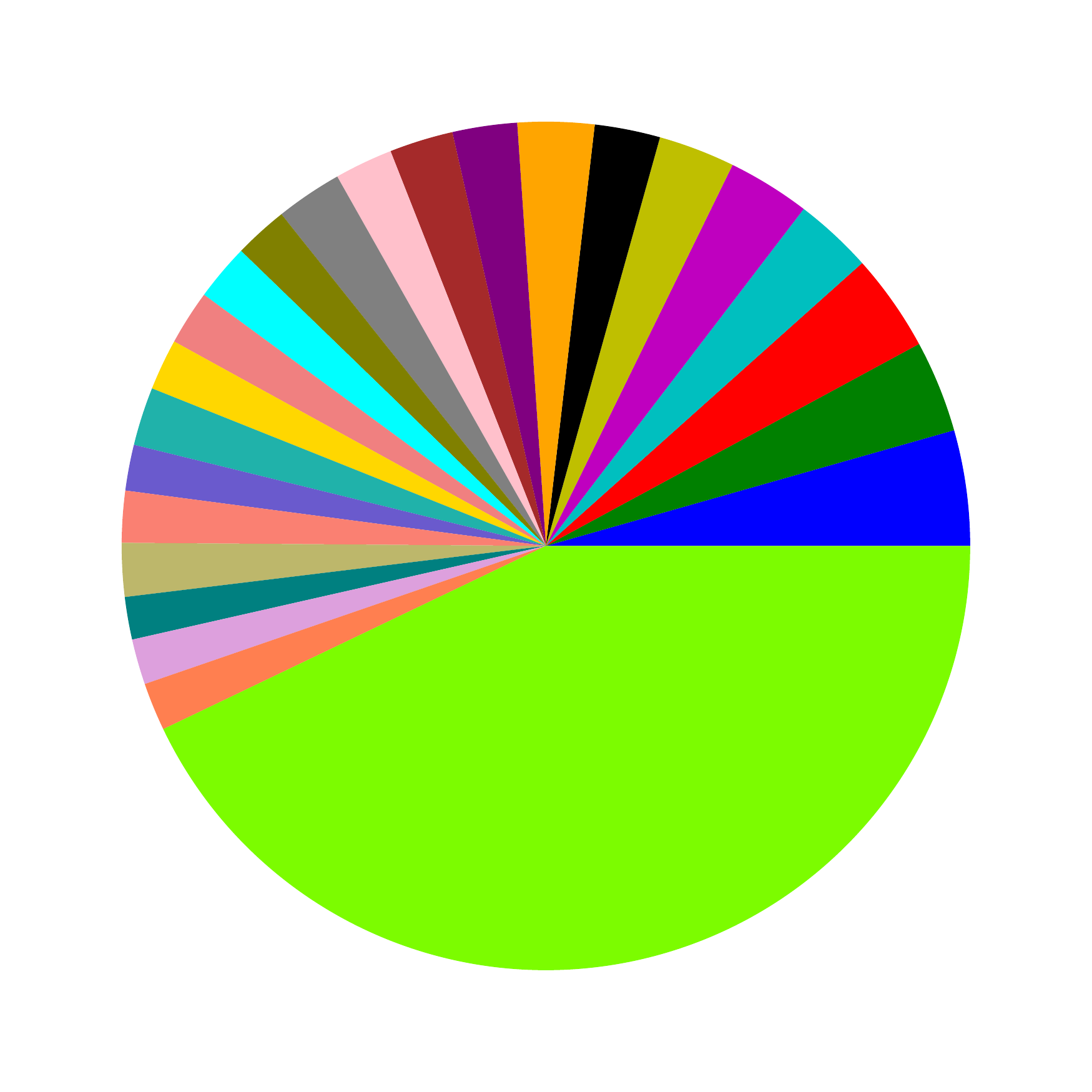}
		\hspace{0.05\columnwidth}
		\includegraphics[trim={0in 0.7in 0in 0.0in}, width=0.38\linewidth]{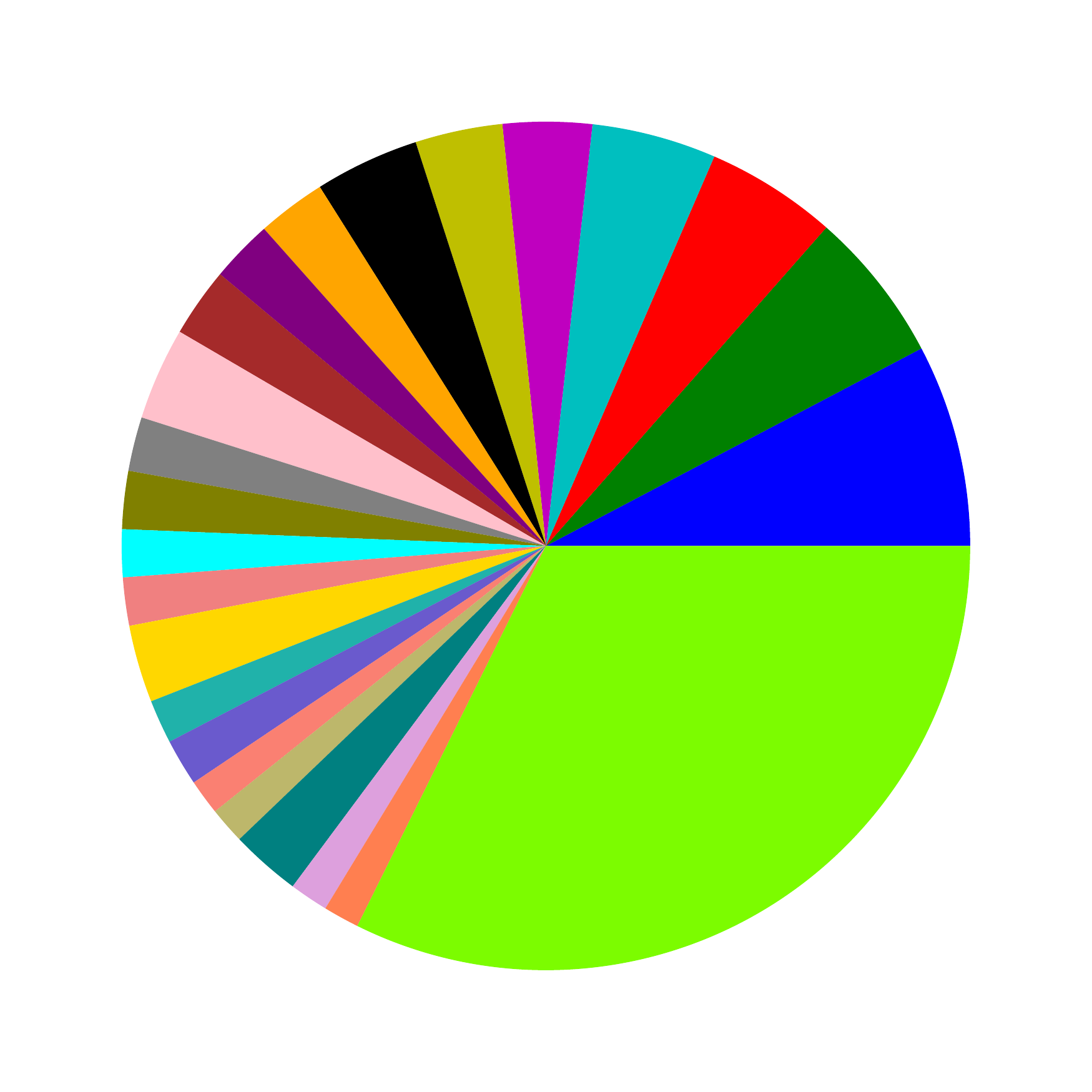}
		\caption{cVAE-GAN errors in WL (left) and BL (right) \\ 989565 total errors observed at $\text{PL}=0$}
	\end{subfigure}
	\caption{Pie charts visualizing error-causing probabilities according to  measured voltages and cVAE-GAN generated voltages at 7000 P/E cycles. The sector labeled {\bf others} combines 41 lesser  frequency patterns. The first column of pie charts correspond to the WL direction and the second column to the BL direction.}
	\vspace{-3ex}
	\label{fig::allcomppie}
\end{figure}

The probabilities for measured and regenerated data are visualized as pie charts in Fig.~\ref{fig::allcomppie}. When we program an interior cell to level 0, there are 64 such patterns of program levels for the pair of adjacent cells in both WL and BL directions. Without ICI effects, the errors would occur randomly for all possible patterns. 

In the measured data, the 23 listed patterns account for 55\% of the errors in the WL direction and around 70\% of the errors in the BL direction. The dominant error pattern in both WL and BL directions is 707. Comparing the area of pattern 707 in WL and BL, we find that pattern 707 in the WL direction is less severe than that in the BL direction.

As shown in Fig.~\ref{fig::allcomppie}, for the prevalent error patterns at 7000 P/E cycles, probabilities observed in the data generated by cVAE-GAN are very similar to those seen in the measured data. The only substantial discrepancy we observed is that the generative model underestimates the fraction of the 707 pattern in the WL direction. At 4000 (resp., 10000) P/E cycles, the model underestimates (resp., overestimates) the  fraction of the 707 pattern in both directions. 
However, at all P/E cycles, cVAE-GAN generates the same rank  ordering of pattern fractions as the measured data in both directions.   

These results show that cVAE-GAN generally produces spatial distributions of voltage levels in a flash memory block that capture  with good accuracy the effects of ICI in both vertical and horizontal directions.

\vspace{-1ex}
\section{Conclusion}
\label{sec::conclusion}

In this paper, we explored the use of conditional generative networks to model the flash memory channel. Unlike traditional modeling and analysis, our model can generate ``realistic'' soft voltage levels from program levels, not only representing accurate statistical distributions among different P/E cycle counts but also preserving spatial ICI effects. Our generative model marks a first step in capturing the spatio-temporal characteristics of flash devices. 


\vspace{-2ex}
\section*{Acknowledgment}
The authors would like to thank Zachary Blair and Yi Liu for the flash memory test platform used in this study. The authors would also like to acknowledge very helpful discussions with  Naoaki Kokubun, Sarah Ekaireb, and Shengqiu Jin.

\vspace{-1ex}
\begingroup

\endgroup
	
\end{document}